\documentclass[conference,10pt]{IEEEtran}
\IEEEoverridecommandlockouts
\usepackage{cite}
\usepackage{xcolor,colortbl}

\def\BibTeX{{\rm B\kern-.05em{\sc i\kern-.025em b}\kern-.08em
		T\kern-.1667em\lower.7ex\hbox{E}\kern-.125emX}}

\usepackage[utf8]{inputenc}
\usepackage[T1]{fontenc}
\usepackage{amsmath,amssymb,amsfonts}
\usepackage[abbreviations]{foreign}
\usepackage{algorithm}
\usepackage{algpseudocode}
\usepackage{graphicx}
\usepackage[margin=3pt,skip=3pt,belowskip=-10pt,font={small,stretch=0.9},labelfont=bf]{caption}

\usepackage[compact]{titlesec}
\usepackage{microtype}
\usepackage{setspace}
\usepackage{listings}
\usepackage{multirow}
\usepackage{scrextend}
\usepackage{subcaption}
\usepackage{textcomp}
\usepackage{booktabs}
\usepackage{xspace}
\usepackage{url}
\usepackage{pifont}% http://ctan.org/pkg/pifont
\usepackage[colorlinks=true,linkcolor=blue,anchorcolor=blue,citecolor=blue,filecolor=black,menucolor=black,runcolor=black,urlcolor=blue]{hyperref}
\usepackage{lipsum}

\usepackage[inline]{enumitem}
\setlist{noitemsep,topsep=0pt,parsep=0pt,partopsep=0pt}

\usepackage[capitalise]{cleveref}

\usepackage{tikz}

\usepackage{multicol}
\definecolor{lightcolor}{rgb}{0,0.5,1}
\usepackage{fontawesome5}
\newboolean{showcomments}
\setboolean{showcomments}{true}
\ifthenelse{\boolean{showcomments}}
{ \newcommand{\mynote}[3]{
		\fbox{\bfseries\sffamily\scriptsize#1}
		{\small$\blacktriangleright$\textsf{\emph{\color{#3}{#2}}}$\blacktriangleleft$}}}
{ \newcommand{\mynote}[3]{}}

\definecolor{darkgreen}{rgb}{0.3,0.5,0.3}
\definecolor{darkblue}{rgb}{0.3,0.3,0.5}
\definecolor{darkred}{rgb}{0.5,0.3,0.3}

\lstdefinestyle{CStyle}{
	backgroundcolor=\color{backgroundColour},   
	commentstyle=\color{mGreen},
	keywordstyle=\color{magenta},
	numberstyle=\tiny\color{mGray},
	stringstyle=\color{mPurple},
	basicstyle=\footnotesize,
	breakatwhitespace=false,         
	breaklines=true,                 
	captionpos=b,                    
	keepspaces=true,                 
	numbers=left,                    
	numbersep=5pt,                  
	showspaces=false,                
	showstringspaces=false,
	showtabs=false,                  
	tabsize=2,
	language=C
}

\newcounter{numobserv} 
\definecolor{beaublue}{rgb}{0.88, 0.93, 0.93}
\usepackage{tcolorbox}
\colorlet{shadecolor}{beaublue}
\tcbset{
	colback=beaublue,
	boxrule=0.5pt,
	boxsep=0pt,
	left=1pt,
	right=1pt,
	top=1pt,
	bottom=1pt,
	sharp corners,
	colframe=white,
}

\title{PhD Forum: Efficient Privacy-Preserving Processing via Memory-Centric Computing}
\author{
	\IEEEauthorblockN{Mpoki Mwaisela}
	\IEEEauthorblockA{\textit{University of Neuchâtel}\\
		Neuchâtel, Switzerland \\
		mpoki.mwaisela@unine.ch}
}

\begin{document}

\maketitle

%!TEX root = main.tex
\begin{abstract} 
	Privacy-preserving computation techniques like homomorphic encryption (HE) and secure multi-party computation (SMPC) enhance data security by enabling processing on encrypted data.
	However, the significant computational and CPU-DRAM data movement overhead resulting from the underlying cryptographic algorithms impedes the adoption of these techniques in practice.
	Existing approaches focus on improving computational overhead using specialized hardware like GPUs and FPGAs, but these methods still suffer from the same processor-DRAM overhead.
	Novel hardware technologies that support in-memory processing have the potential to address this problem.
	Memory-centric computing, or processing-in-memory (PIM), brings computation closer to data by introducing low-power processors called \emph{data processing units} (DPUs) into memory.
	Besides its in-memory computation capability, PIM provides extensive parallelism, resulting in significant performance improvement over state-of-the-art approaches.
	We propose a framework that uses recently available PIM hardware to achieve efficient privacy-preserving computation.
	Our design consists of a four-layer architecture:
	(1) an \emph{application layer} that decouples privacy-preserving applications from the underlying protocols and hardware;
	(2) a \emph{protocol layer} that implements existing secure computation protocols (HE and MPC);
	(3) a \emph{data orchestration layer} that leverages \emph{data compression} techniques to mitigate the data transfer overhead between DPUs and host memory;
	(4) a \emph{computation layer} which implements DPU kernels on which secure computation algorithms are built.
	
\end{abstract}
%!TEX root = main.tex
\section{Problem definition and planned contribution}
Privacy-preserving computation (PPC)~\cite{ppc} enables processing operations on data without revealing the data to untrusted parties.
PPC techniques can be cryptographic-based, \ie homomorphic encryption (HE)~\cite{he} and secure multi-party computation (SMPC)~\cite{smpc} , both referred to as \emph{secure computation} (SC), or hardware-based, \eg via trusted execution environments (TEEs)~\cite{tee}.
We focus on HE and SMPC.
Homomorphic encryption enables computations to be performed directly on encrypted data.
This enables clients to securely outsource sensitive data and computation to an untrusted cloud.
Secure multi-party computation (SMPC) on the other hand enables multiple (distrusting) parties to collaboratively compute a function $f$ on their sensitive data $x_1,\ldots, x_n$ without revealing information about each party's private data other than what can be inferred from the output $f(x_1,\ldots, x_n)$.
%Common applications of privacy-preserving computation are in the health~\cite{xx} and financial~\cite{bibid} industries.

Besides their very high computation overhead, SC techniques like HE and SMPC introduce high memory overhead~\cite{mage}.
This is because common SC workloads like privacy-preserving machine learning (PPML) and secure data analytics deal with large amounts of data.
Moreover, the \emph{data amplification} inherent in SC approaches further increases the memory footprint of these applications.
For example, a 64-bit plaintext integer can take up to 1KB of memory when using a garbled circuit (an SMPC protocol)~\cite{mage}.
This rapid growth in ciphertext sizes leads to poor data locality, increasing data movement between the processor (\ie CPU) and memory.
The \emph{limited bandwidth} of the memory channel introduces a significant performance bottleneck for these workloads.
Prior research works have focused mainly on addressing the computational overhead of SC using accelerators like GPUs~\cite{piranha, wang12, jung21} or FPGAs~\cite{cousins17}. %However, the data movement overhead is inherent in these approaches as well.

\smallskip\noindent\textbf{Planned contribution. }
The data movement bottleneck can be addressed by shifting from processor-centric designs towards more memory-centric designs.
Processing-in memory (PIM) architectures augment memory with compute capabilities, thereby mitigating the memory bandwidth bottleneck.
We focus on recent on-the-market PIM technology provided by UPMEM~\cite{upmem-pim}.
UPMEM's PIM architecture introduces general purpose processors called \emph{DRAM processing units} (DPUs) into memory.
Every $64$MB chunk of memory in a UPMEM DIMM is augmented with a DPU, allowing computing capability to scale with memory size~\cite{nider21}.
Furthermore, each DPU supports up to $24$ threads (called \emph{tasklets}) capable of executing code concurrently.
This provides extensive parallelization capabilities absent in CPU-based designs.
Recent works leverage PIM architectures to accelerate machine learning~\cite{rhyner2024AnalysisOD}, generic computing algorithms~\cite{nider21}.
Unfortunately, there is a lack of frameworks that leverage PIM to accelerate secure computation.
In this context, we seek to \textit{build an efficient and scalable privacy-preserving computation framework based on PIM}.

\smallskip\noindent\textbf{Roadmap.} The remainder of this paper is organized as follows:
Section \S\ref{sec:solution} presents the research methodology and approach, detailing the methods and strategies employed in this study.
Section \S\ref{sec:rw} reviews the related work, providing an overview of existing research and how it compares to our approach.
Finally, Section \S\ref{sec:conclusion} concludes the paper, summarizing the key findings and discussing potential directions for future research.

\section{Research methodology and approach}
\label{sec:solution}

\begin{figure}[!t]
	\vspace{-7mm}
	\centering
	\includegraphics[scale=0.6]{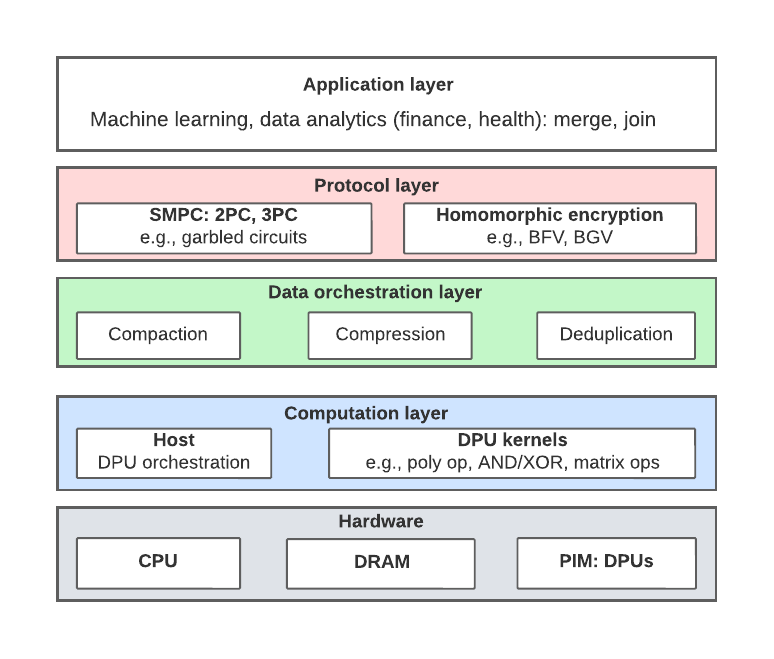}
	\caption{The proposed system features a hierarchical architecture, spanning from the hardware layer (CPU, DRAM, and PIM) to the application layer (machine learning and data analytics).}
	\label{fig:arch}
	% \vspace{7mm} 
\end{figure}

In contrast to prior research that focus on accelerating specific secure computation techniques~\cite{saransh21, saransh24, Giannoula2022SparsePPomacs} or specific algorithms~\cite{gilbert24}, this paper proposes a generic and extensible framework for PIM-based acceleration of secure computation.
The proposed framework follows a layered architecture (inspired by ~\cite{piranha}), with each layer containing separate modules that employ different techniques to address specific problems assigned to that layer.
The modular architecture of \autoref{fig:arch} enables for seamless integration of privacy-preserving approaches into other domains without modifying the basic application logic.

\subsection{Application layer}
The Application Layer is where specific use cases are developed, such as machine learning applications and data analytics.
This layer utilizes the services and capabilities provided by the underlying Protocol Layer, allowing developers to build and deploy privacy-preserving solutions.
It focuses on implementing domain-specific functionality while relying on the security mechanisms managed by the layer below.

\subsection{Protocol layer}
The purpose of this layer is to employ protocols and schemes that ensure privacy-preserving processing.
This includes cryptographic protocols like HE ~\cite{bgv2012} schemes such as BFV (Brakerski-Fan-Vercauteren) and BGV (Brakerski-Gentry-Vaikuntanathan), which enable computation on encrypted data, as well as SMPC protocols relying on either secret-sharing ~\cite{beimel2011secret} or garbled circuits~\cite{yao-gc}, which could involve two (2PC), three (3PC), or more parties.

\subsection{Data orchestration layer}
%While PIM hardware provides advantages in terms of extensive parallelism and improved memory bandwidth, 
The processing-in-memory model~\cite{pim_model} typically comprises three components:
\textbf{CPU-side} with parallel cores and fast access to a small shared memory;
\textbf{PIM-side} consisting of multiple PIM modules, each with cores, i.e., DPUs, with a local memory;
and \textbf{Network} connecting the CPU-side and PIM-side.
Therefore, this layer aims to handle the data representation for efficient data transfer between CPU-side and PIM-side.

The UPMEM-PIM design separates physical address spaces for PIM and CPU-side or host memory ~\cite{lee2024}, requiring data to be explicitly copied between them.
This overhead dominates the overall cost of offloading operations to DPUs.
A simple element-wise vector addition benchmark is presented to understand this overhead.
\begin{figure}[!t]
	\vspace{-6mm}
	\centering
	\includegraphics[scale=1.1]{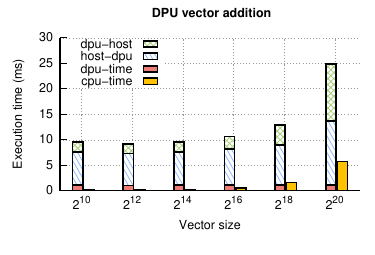}
	\caption{Data movement overhead for vector addition on DPUs.}
	\label{fig:vector-add}
	\vspace{4mm} 
\end{figure}
\autoref{fig:vector-add} outlines the cost of copying the input data from the host to the DPUs (\texttt{host-dpu}), the overall processing time, \ie addition operations on the input vectors (dpu-time), and the cost of copying the results from the DPUs back to the host (\texttt{dpu-host}).
This is contrasted against a CPU-only baseline (\texttt{cpu-time}).
We observe that data copy operations from host to DPU and vice versa account for up to $51\%$ and $45\%$ of the overall cost for the DPU-based benchmark.
These results are in line with recent studies that characterize DPU data transfer overhead~\cite{lee2024}.

Motivated by our performance analysis of the data copy overhead between host and DPUs, we introduce an approach to mitigate the data copy overhead between host memory and DPU memory by reducing the data copied.
We achieve this by leveraging the following techniques: (1) \emph{lightweight byte-oriented compression} and (2) \emph{data deduplication}.

Algorithm~\ref{algo:dmc} summarizes the overall workflow of the data orchestration layer.
A dataset (on the host) of size $N_D$ is split evenly amongst $N$ DPUs, each of which runs $T$ tasklets.
To mitigate the overhead of host-DPU copy operations, this data is compressed on the host side, copied to the DPUs, and then decompressed, after which processing operations are performed on the data.
Similarly, the results are compressed, and copied back to the host.

\begin{algorithm}[!t]
	\small
	\caption{Data orchestration layer}
	\begin{algorithmic}[1]
		\State $N$: number of DPUs
		\State $N_D$: data size per DPU
		\State $T$: number of tasklets per DPU
		\State Split data into $N$ chunks of equal size $N_D$
		\For{$i \gets 0$ \textbf{to} $(N-1)$}
		\State {Compress}(chunk-i)
		\State Copy compressed chunk-ito DPU$_i$'s MRAM
		\State {Decompress}(chunk-i) with $T$ tasklets
		\State Process chunk-i on DPU$_i$ with $T$ tasklets
		\State {Compress} results from chunk-i
		\State Copy compressed results from DPU$_i$ to host
		\State {Decompress} results from chunk-i
		\EndFor
		\State Host aggregates decompressed results from all DPUs
	\end{algorithmic}
	\label{algo:dmc}
\end{algorithm}

We propose the following data compression techniques:
%\noindent\textbf{Bit packing.} This is a technique used to reduce the amount of space required to store fix-side data elements by packing them more efficiently.

\noindent\textbf{Byte-oriented compression.}
The compression method must be simple enough to be implemented on a DPU and fast enough so a compression + copy + decompression cycle for a data block is faster than copying the uncompressed data block between the DPU’s memory and host DRAM.
Byte-oriented compression algorithms like Vbyte~\cite{vbyte} are well suited for this purpose, especially for integer arrays.
The basic idea behind byte-oriented compression is that not all integers require 4 bytes for storage.
For example, integers in the range $[0,2^2)$ can be stored with a single byte, instead of 4, resulting in a compression ratio\footnote{Data compression ration = uncompressed size/compressed size} of up to 4.
Generally, HE and SMPC schemes primarily operate on integer kernels. Typically, represented as vectors, often corresponding to polynomial formats ~\cite{piranha,ckks,he,smpc}.
This makes byte-oriented compression a good candidate for this usecase, as it can effectively reduce the size of integer-based vectors and optimize data transfer efficiency in secure computation algorithms.

\begin{figure}[!t]
	% \vspace{3mm}
	\centering
	\includegraphics[scale=1.1]{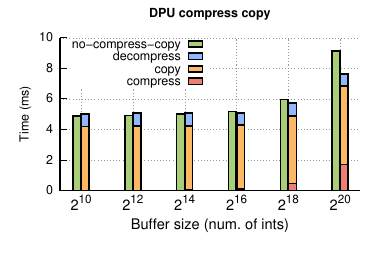}
	\caption{DPU byte-oriented compression for integers in the range $[0,2^2]$.}
	\label{fig:vbyte_compression}
	%\vspace{-6mm} 
\end{figure}

\autoref{fig:vbyte_compression} demonstrates preliminary performance evaluation of the compression function proposed.
For polynomials with coefficients in the range $[0,2^2)$ byte-oriented compression achieves a data compression ratio of up to $\approx1.34\times$ for an unsigned integer array of size $2^{20}$.
This compression ratio can be improved if the array is sorted~\cite{vbyte}.
This suggests data compression can indeed be leveraged to mitigate the host-DPU copy overhead.

%\noindent\textbf{Dynamic Markov compression.} Contrary to data compaction, \eg via bit packing, data compression is a technique that exploits patterns and redundancy to reduce the size of data.

\noindent\textbf{Data deduplication.} This is a technique that eliminates duplicate copies of repeating data so as to reduce data size.
Similarly to byte-oriented compression, dedpulication techniques can be explored as a possible solution to the data copy problem between host memory and DPU memory.
A key challenge is ensuring that blocks of compressed data remain self-contained (\ie they do not depend on blocks present in other DPUs) after they are copied.

\subsection{Computation layer}
The computation layer implements DPU kernels (\eg polynomial operations for plaintext and ciphertext vectors), and handles DPU orchestration on the host side, \eg launching DPUs and copying data to DPU memory.

%!TEX root = main.tex
\section{Related work}
\label{sec:rw}
%\vspace{0.25cm}
We classify the related work into three categories:
\emph{(i)}~systems that leverage PIM architectures for accelerating privacy-preserving computation,
\emph{(ii)}~systems that leverage PIM architectures for accelerating generic applications, and
\emph{(iii)}~systems that leverage other hardware architectures for accelerating privacy-preserving computation.

\looseness-1
\smallskip\noindent\textbf{PIM-based acceleration for privacy-preserving computation.}
Gupta \textit{et al.}~\cite{saransh21} conducted a theoretical analysis of the performance impact of PIM in accelerating homomorphic encryption algorithms.
MemFHE~\cite{saransh24}, CryptoPIM~\cite{DBLP:conf/dac/NejatollahiGIRC20} and Jonatan \textit{et al.}~\cite{gilbert24} evaluated the impact of PIM on the NTT algorithm, commonly used in fully homomorphic encryption libraries for efficient polynomial multiplication.

While these works extend the state-of-the-art on secure computation with memory-centric computing, they focus on a single secure computation technique, \ie homomorphic encryption, and do not provide techniques to address key issues like data copying in PIM-based designs.
Our design provides a more holistic and generic framework, with a modular architecture that can be integrated across a wide range of secure computation problems.

\smallskip\noindent\textbf{PIM-based acceleration for generic applications.}
Nider et al~\cite{nider21} evaluate the impact of UPMEM PIM across various commonly used algorithms, and emphasize on the advantages of PIM-based design in terms of performance as compared to CPU-centric designs.
Other recent works have explored PIM-based acceleration for specific use cases like matrix-vector multiplication~\cite{Giannoula2022SparsePPomacs}, DNA sequencing~\cite{DBLP:conf/bibm/LavenierRF16}, or image decoding~\cite{DBLP:conf/systor/NiderDGNF22}.

Our research is aligned with these studies but focuses more on privacy-preserving computation.

\smallskip\noindent\textbf{Non PIM-based hardware acceleration for privacy-preserving computation.}
Over the past decades, numerous studies have explored hardware accelerators like GPUs and FPGAs to accelerate secure computation. 
Piranha~\cite{piranha} proposes a general-purpose, modular platform for accelerating secret sharing-based MPC protocols using GPUs.
Various research works~\cite{wang12, DBLP:conf/iscas/WangCH14, DBLP:journals/tc/WangHCHS15, DBLP:journals/tetc/BadawiPAVR21, shivdikar23} have leveraged GPUs to accelerate fully homomorphic encryption algorithms.
While these GPU-based approaches generally outperform traditional CPU methods, they still necessitate transferring data from main memory to dedicated GPU memory, a drawback that PIM technology aims to address.

On the other hand, some works like~\cite{feldmann21, sinha19, DBLP:journals/tc/TuranRV20, DBLP:conf/asplos/RiaziLPD20} leverage FPGAs to accelerate homomorphic encryption.
Similarly, these FPGA-based solutions face challenges with high data movement, which PIM-based solutions address.
% ---

\section{Conclusion}
\label{sec:conclusion}
This study presents a memory-centric privacy-preserving processing framework using a layered architecture from hardware components to application development. 
It encapsulates secure computation schemes like homomorphic encryption and secure multiparty computation to enforce privacy. 
The approach optimizes overhead introduced by data copy using compression and deduplication strategies. 
Future work will explore scalability and application to privacy-sensitive domains.
%!TEX root = main.tex
\section*{Acknowledgment}
This work was supported by the Swiss National Science Foundation under project P4: Practical Privacy-Preserving Processing (no. 215216).%(no. 200020\_215216).

\bibliographystyle{plain}
\bibliography{fhe-pim, mpc, min}

\end{document}